\documentclass[runningheads]{llncs}
\usepackage{graphicx}
\usepackage{algorithm}
\usepackage[noend]{algpseudocode}
\algnewcommand\algorithmicforeach{\textbf{for each}}
\algdef{S}[FOR]{ForEach}[1]{\algorithmicforeach\ #1\ \algorithmicdo}

\newcommand{\circled}[1]{\raisebox{.5pt}{\textcircled{\raisebox{-.9pt} {#1}}}}
\newcommand{\anovatest}[4]{\textit{F(#1)} = #2; $p<$#3; $\eta^2_p$=#4}
\newcommand{\msd}[2]{\textit{M}=#1; \textit{SD}=#2}
\usepackage{xfrac}
\usepackage{macros}
\usepackage{enumitem}
\usepackage{amsmath}
\usepackage{comment}

\begin{document}
\title{Immersive and Interactive Visualization of 3D Spatio-Temporal Data using a Space Time Hypercube Applied to Biological Imaging Data}%\thanks{Supported by organization x.}}
\titlerunning{Immersive Space-Time Hypercube Visualization for Biological Imaging}

\author{Gwendal Fouch\'e \inst{1} \and
Ferran Argelaguet\inst{1} \and
Emmanuel Faure\inst{2} \and
Charles Kervrann\inst{3}}
\authorrunning{G. Fouch\'e et al.}
\institute{Inria, Univ Rennes, IRISA, CNRS, Rennes, France.\and
LIRMM, Univ Montpellier, CNRS, Montpellier, France. \and
Inria Centre Rennes-Bretagne Atlantique, UMR144 CNRS Institut Curie, PSL Research University, Sorbonne Universités, Paris, France
}

\maketitle              % typeset the header of the contribution
\begin{abstract}
%%%
    We propose an extension of the well-known Space-Time Cube (STC) visualization technique in order to visualize time-varying 3D spatial data, taking advantage of the interaction capabilities of Virtual Reality (VR). 
    The analysis of multidimensional time-varying datasets, which size grows as recording and simulating techniques advance, faces challenges on the representation and visualization of dense data, as well as on the study of temporal variations.
    First, we propose the Space-Time Hypercube (STH) as an abstraction for 3D temporal data, extended from the STC concept.
    Second, through the example of embryo development imaging dataset, we detail the construction and visualization of a STC based on a user-driven projection of the spatial and temporal information.
    This projection yields a 3D STC visualization, which can also encode additional numerical and categorical data.  
    Additionally, we propose a set of tools allowing the user to filter and manipulate the 3D STC which benefits from the visualization, exploration and interaction possibilities offered by VR.
    Finally, we evaluated the proposed visualization method in the context of the visualization of spatio-temporal biological data. 
    Several biology experts accompanied the application design to provide insight on how the STC visualization could be used to explore such data.
    We report a user study (n=12) using non-expert users performing a set of exploration and query tasks to evaluate the system. 
    %The second user study (n=5) report the feedback from five biologists\added{, who also accompanied the application design as consultants,} providing insights on how the STC visualization could be used for the exploration of complex spatial-temporal data.
    %mention their implication in the design?
%%%%
\end{abstract}
\begin{figure*}[h!]

  \includegraphics[width=\textwidth]{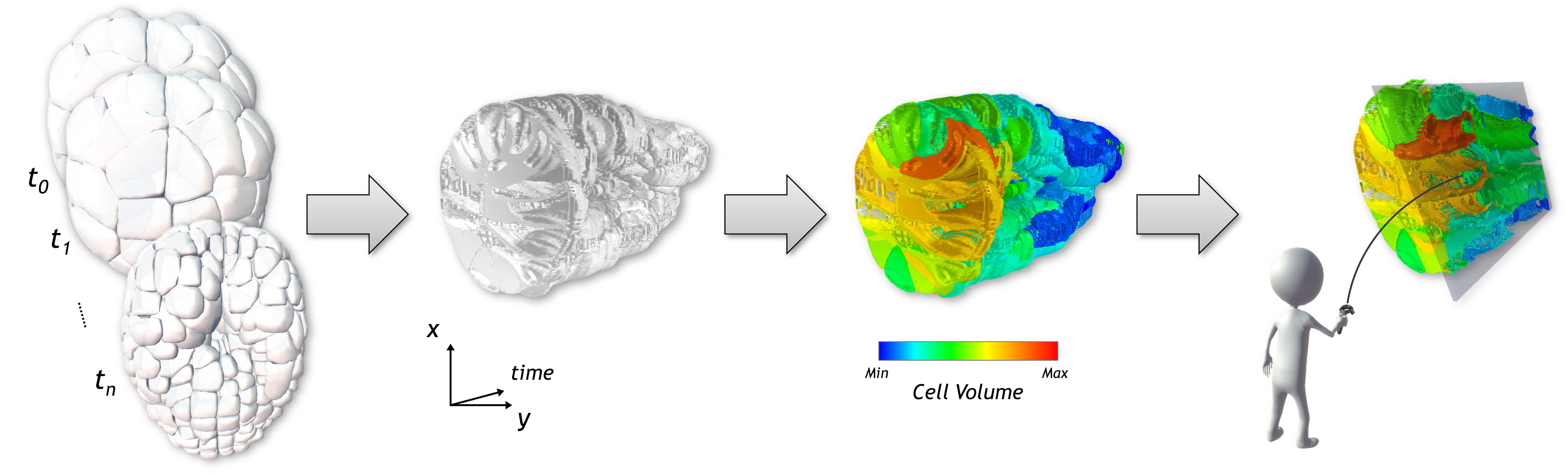}
  
  \caption{From a cross-section on the 3D surface-based temporal data shown on the first figure, we generate a Space-Time Cube visualization, displayed in the second image, showing the evolution over time of the spatial data of the cross-section displayed on the x and y axes. The third picture shows how the visualization can be enriched with quantitative and qualitative data using different color coding. A set of interaction tools help the user to explore the generated visualization, as seen in the last image.
  %This last sentence is not illustrate in the figure
  } 
  \label{fig:teaser}
\end{figure*}

%\tableofcontents
\section{Introduction}

Immersive Analytics is an emerging research topic which ``investigates how new interaction and display technologies can be used to support analytical reasoning. [...] Immersive Analytics builds on technologies such as large touch surfaces, immersive Virtual and Augmented Reality environments, haptic and audio displays, and modern fabrication techniques''~\cite{chandler2015immersive}. Immersive analytics, as reported by Skarbez et al. \cite{skarbez2019immersive}, leverages large field of view displays, stereoscopic rendering and rich spatial 3D interactions in order to increase spatial understanding, decrease information clutter and help with visual analytics tasks.
In particular, immersive 3D environments can help with flexible exploration and interaction with complex time-varying 3D spatial data. For example, the enhanced depth perception offered by stereoscopy in Virtual Reality improves the visualization of details in volumetric data~\cite{laha2012effects}, while, having a large workspace allows for the juxtaposition of multiple coordinated and interactive views~\cite{johnson_bento_2019}. 

Methods from the field of Immersive Analytics could notably be beneficial for biological imaging, where large time-varying datasets are more and more present, and have emerging constraints from the growing spatio-temporal resolutions. For instance, Lattice Light-Sheet Microscopy\cite{chen2014lattice} allows the capture of volumes at a resolution of a few hundred nanometers at a few seconds interval. This technology is notably used for recording embryo development\cite{guignard2017contact}, creating terabytes of raw data, in contexts of analysis in embryology, morphodynamics, and intracellular dynamics. 
%
%//TAG : describe data
In such context, extracting surface meshes to describe volumes (see Figure~\ref{fig:teaser}, left) is a commonly used solution in order to display and analyze data, as proposes the web-based application MorphoNet~\cite{leggio2019morphonet}.
The output dataset is composed a group of objects defined by their shapes, that evolve over time.
%
%Depending on the entity imaged, the visualization is more or less dense, and the tracking of object can be quite well defined thanks to the high temporal resolution.
%
Moreover, spatio-temporal numerical and categorical data are often available for these datasets, for example, cell information (e.g. volume, lifespan, name, genetic information) can be available for each time step. Thus, visualization and interaction for analyzing this type of time-varying multidimensional datasets become challenging as well. 
Widely used software in biology imaging, such as ImageJ~\cite{rasband1997imagej} or Imaris~\cite{Imaris}, only propose very basic non-interactive 3D viewers.

%Among the software most widely used by biology experts, ImageJ~\cite{rasband1997imagej}, an open source image analysis tool, proposes a 2D temporal interactive interface as well as a non-interactive 3D visualizer, which handles very basic animation functionalities for 3D temporal data.
%
%Imaris \cite{Imaris} includes a viewer for 3D and 4D data, with temporal exploration based on user-defined events.
%
%These tools are not designed for the interactive exploration of 3D temporal datasets.

%This can be an efficient method to display complex encoding of multidimensional datasets, under constraints of occlusion and readability, especially in case of time-varying data.

%With the emergence of recent accessible VR hardware such as the HTC Vive, the challenge of integrating such tools in analysts' workflow seems more and more achievable. 

%
%\gwendal{These features led us to consider adapting usual visualization techniques to 3D temporal data, aided by immersive environment.}

%\ferran{novelty with respect to previous works, what is new with respect the star. }

%Propose STH which is an extension of the STC
%Novelty
%Temporal and spatial continuity. -> justified as it's an important criterion for a decent analysis of dynamic features

%Evaluation - Comprehension of the visualization STH, complex 3D structure -> naive participants with the goal to explore whether they were capable of understanding the visualization and use it to extract spatial and temporal events.
The main objective of this paper was to propose visualization methods enabling the comparison of 3D spatio-temporal data in a more direct way. 
For this purpose, we propose an extension of the Space-Time Cube (STC) visualization technique~\cite{bach_review_2014} to 4D spatio-temporal data, named hereinafter Space-Time Hypercube (STH). 
Conceptually, the STH considers that the data to visualize lays in a 4D hypercube with three spatial dimensions and one temporal dimension. 
To enable the direct visualization of the 4D hypercube, we extended the classical operations of the STC to generate varied meaningful 3D visualizations, that are juxtaposed to obtain a compact overview of spatio-temporal data.
As an example, we propose a projection operation on the hypercube on datasets of embryo development 3D temporal imagery.
The method is independent of the type of spatial representation (e.g. mesh, volumetric), and projects the hypercube into a 3D volume that can be directly visualized.
The projection operation relies on a user-defined cross-section on the spatial dimension. 
This cross-section is computed along the temporal dimension and stacked into a 3D volume, hence a STC, which can be enriched with color mapped data (see Figure.~\ref{fig:teaser}). 

Due to the complexity of the visualized structures and the need for 3D interaction methods to define the projection operation, we proposed the use of this method in an immersive context to leverage the benefits of immersive analytics. 
We expected that the enhanced depth perception would improve the visual extraction of meaningful structures of the generated STC, and the use of 3D user interfaces would allow users to explore naturally the STC from different perspectives, as suggested in several works, in general cases~\cite{loomis2003visual} and specifically for STCs~\cite{bach_descriptive_2017,ssin_geogate:_2019,wagner2019evaluating}. 
Furthermore, the enlarged interaction space of VR allows the juxtaposition of the STC and snapshots of the original spatio-temporal data, enabling the synchronized exploration and manipulation of the two visualizations.
In order to provide adequate tools for basic analysis of our use case data, we considered the input from domain experts to improve the design of the application.
Finally, the paper presents a use-case illustrating the usages of the STH to generate meaningful visualization of spatio-temporal data in embryo developmental studies, as well as an evaluation of this method with non-expert users.
%both with non-expert and domain expert users.
%
%The STH approach opens a novel perspective for the visualization and the interaction for spatio-temporal data

In summary, the main contributions of the paper are:

\begin{itemize}
%    \item \removeKeep{A novel visualization method for spatial 3D temporal data based on the Space-Time Cube method, with a generation algorithm independent of the spatial data encoding.}
%    \item \removeKeep{Description of VR system for the visualization of the STC including specific 3D interaction methods for the exploration and manipulation of the juxtaposed views.}
    \item An extension of the well-known Space-Time Cube technique as an abstraction of 3D temporal data, named Space-Time Hypercube;
    \item A projection operator of the 4D STH to generate a 3D STC that can be visualized in an Immersive Analytics application;
    \item A formal evaluation of the proposed method on embryology use cases.
\end{itemize}

%The remainder of the paper is structured as follows. Section \ref{relatedworks} will present an overview of related works on temporal visualization, Space-Time cubes and immersive analytics frameworks. Section \ref{STCsection} will explain more thoroughly our works on the Space-Time cube visualization and interaction. Then, we will give in Section \ref{usecases} examples of use cases for our tool in a biological imaging context, followed in Section \ref{evaluation} by an evaluation of our method in this context by non-expert users. Finally, in Sections \ref{feedback} and \ref{discuss} we will discuss the strengths and weaknesses of our work with expert feedback and provide concluding remarks.

\section{Related Work}\label{relatedworks}
%We explored various methods for visualization of time varying data. We put a focus on a generic method, the Space-Time Cube, as well as existing immersive frameworks for visualization of multidimensional data. We will present the most relevant works in this section.}

Time-varying data are more and more present in scientific visualization, and numerous methods have been developed to visualize the time component and space-time relationships.
Such methods can focus on the visualization of point clouds~\cite{li2013analyzing,lu2008interactive}, more~\cite{neuroth2019interactive,huang2015trajgraph,liu2016smartadp,hurter2009fromdady} or less~\cite{amirkhanov2019manylands,andrienko2017clustering} large trajectory datasets, simulated flow data~\cite{waser2010world,schindler2012multiverse} and surface or volumetric datasets~\cite{kim_comparison_2017}. Immersive visualization methods are also becoming commonplace~\cite{hurter2018fiberclay,homps2020revivd}.
However, due to the vast literature in this domain, we constrain the state of the art to visualization methods adapted to datasets representing time-varying 3D spatial data with a particular focus on Space-Time Cube visualizations methods, which is the main scope of the paper.
For a comprehensive review of spatio-temporal visualization methods, we refer the reader to the following comprehensive works ~\cite{aigner2011visualization,andrienko2013visual,bai2020time}.

\subsection{Visualization of Time-varying Volume Data}
    
%Time-varying data are more and more present in scientific visualization, and several methods have been developed to visualize the time component and space-time relationships.
%
%. Yet, Kim et al.~\cite{kim_comparison_2017} reported a lack of visualization tools for 3D volumetric temporal data, and especially surface based 3D temporal datasets.
%
A number of approaches exist to visualize time-varying data: dynamic methods, implying animation and interaction, and static visualizations of either the full data, or extracted lower-dimension data.
    
\textbf{Dynamic Visualizations.} Animation is an intuitive method to explore time evolution in 3D temporal datasets. 
In designs using animation, time increases automatically, showing the evolution of the data through time. 
%
%The work of Coffey et al.~\cite{coffey_visualizing_2012} showed that animated and interactive design choices for time exploration can decrease the amount of error in analysis tasks. 
%
%They also suggested that hybrid methods that would implement interactive, animated and/or static time exploration designs could compensate for the flaws of other methods. 
%
%For instance, Akiba et al.~\cite{akiba_aniviz:_2010} proposed Aniviz, an interactive animation interface enabling the exploration of  time-varying volumetric data by editing a time-dependent transfer function.
%
%The user can specify display, animation and data related parameters.
%
Several works mention the use of animated and interactive design choices for time exploration~\cite{coffey_visualizing_2012,akiba_aniviz:_2010}, yet animation can be less adequate for comparison tasks~\cite{kim_comparison_2017} or the analysis of space-time-value relationships~\cite{woodring_multi-variate_2006}. 
There is a higher cognitive load for the user who has to remember the state of the data between different moments, which can limit the observation of details, as the slogan ``\textit{Eyes Beat Memory}'' suggests. 
Some methods have been proposed to cope with this issue, such as time-warping in animation~\cite{solteszova2020memento}, yet, a more intuitive solution lies in static visualizations.
%
%Solteszova et al.~\cite{solteszova2020memento} proposed to use time-warping in animations to cope with this issue. 
%Their method allows the user to manipulate time evolution of a video or a selected point of interest in the video. 
%
%Yet, a more intuitive solution for the fore-mentioned issue lies in static visualizations.

\textbf{Static Visualizations.} Woodring and Shen~\cite{woodring_multi-variate_2006} proposed a solution for static temporal visualization for time-varying volumetric data, based on set and numerical operations to filter the displayed information, reduce occlusion and focus on points of interest. A color mapping can be applied to overlay chronological information~\cite{woodring_multi-variate_2006}.
With this method, the temporal evolution is visualized by merging the spatio-temporal data into a volume. 
Although, there is a compromise to make between occlusion and the amount of displayed data, it is less suited for dense data.
Alternatively, in order to reduce the data to display, regions of interest can be defined either manually or automatically. 
For example, Lu and Shen~\cite{lu2008interactive} used a dissimilarity measure to extract single time steps in the volumetric temporal data, eventually displaying only the essential time steps on a color mapped timeline.
%TransGraph, a tool developed by Gu and Wang \cite{gu_transgraph:_2011}, goes even further in the dimension reduction creating a graph of points of interest extracted from blocks of volumetric data. Correspondences between nodes and volume regions are handled by a brushing and linking method. This allows the user to apprehend the evolution of 3D temporal data over time through a 2D interactive graph representation. 
%
%However, Kim et al. \cite{kim_comparison_2017} reports a lack of use of virtual environment for spatial and volumetric data. 
%
In the context of a more recent work by Johnson et al.~\cite{johnson_bento_2019} proposed BentoBox, a Virtual Reality data visualization interface for simulated volumetric and temporal data, which juxtaposes several instances of a volumetric dataset under different parameters, in an array layout.
BentoBox disposes of a range of tools focused on the comparison of datasets and manipulation of parameters, using animations, color mapping and 3D bimanual interactions. 

Finally, a versatile visualization technique which enables the compact visualization and manipulation of time-varying data, is the Space-Time Cube which is detailed in the following section.

\subsection{Space-Time Cube Visualization}

First introduced by Hägerstrand~\cite{hagerstrand69people} as time-space volume and paths in a context of socio-economic study, the Space-Time Cube (\textit{STC}) is a representation using two axes for data and a third axis for time. It is used for numerous representations of time-varying data, would it be for geometrical illustration, geographic data~\cite{ssin_geogate:_2019,leduc2019space} or trajectories\cite{wagner2019evaluating}.
Bach et al.~\cite{bach_review_2014} summarized, in their review of temporal visualizations based on the STC, different types of operations that can be applied to visualize data in a STC.
Cutting operations allow the extraction of an image at a particular moment in time, as short exposure photography, or a particular plane in the data space, a slice of the cube. On contrary, flattening operations collapse the STC along an axis to obtain a 2D representation, when the time is collapsed, it can be depicted as a long exposure photograph. 
In addition, depending on the data and the visualization design, 3D rendering, interpolation, volume extraction, non-orthogonal or non-planar operations can also be used. Later work by the same authors~\cite{bach_descriptive_2017} generalized the notion of STC visualization, modelling various visualization methods as associations of operations on a STC. 

Among the topics emerging from STC visualizations, two kept our attention. First, \textbf{higher dimensions} in the data, such as 3D temporal data, can be visualized as a higher dimension Space-Time Hypercube. In the case of spatio-temporal data, operations of cutting, corresponding to a projection of the 4D data, could yield a visualizable 3D image. 
If the idea of a higher dimension STC was evoked in Bach et al.~\cite{bach_review_2014,bach_descriptive_2017} reviews, this lead has not been thoroughly explored.
Among the works that are the closest to it,
%the Woodring and Shen's~\cite{woodring_multi-variate_2006} method described previously already evoked a time flattening operation on a STH;
%
%in this direction, 
Woodring et al.~\cite{woodring2003high} proposed a method to generate and render 3D slices based on an arbitrary hyperplane equation. 
The authors gave a few guidelines to interpret the output visualization with particular value of the hyperplane equation~\cite{woodring2003high}. 
However, the intuitiveness of the equation parameterization and concrete interpretation of the visualization have not been formally evaluated.
%
%Finally, another method was proposed by Krone et al.~\cite{krone2016molecular} in which the focus was on the evolution of the surface of molecular membranes. In this scenario, surface data for each time step was projected into a plane, enabling the visualization of the evolution of the membrane in a STC.

The second emerging topic from STC visualizations is \textbf{interactivity}. If Bach et al.~\cite{bach_descriptive_2017} advice against operations of 3D rendering of STCs, it seems that interaction and navigation methods, notably with immersive technologies~\cite{bach2017hologram} could help in solving some of the difficulties evoked~\cite{bach_descriptive_2017}, such as occlusion, depth ambiguity and perspective distortion. This would help take advantage of the general overview of the data that this operation can offer.
STC visualization can also take advantage to direct interactions in immersive environments.
%
%For example, the GeoGate system~\cite{ssin_geogate:_2019} displays geographic trajectories (2D + time) in an augmented reality environment using a STC. 
%
%A tabletop system displaying satellite information is coupled with the STC visualization to provide additional context information. 
%
%Furthermore, the user can interact with the data through a tangible ring controller, providing natural interactions. 
%
For instance, Filho et al.~\cite{wagner2019evaluating} implemented a STC displaying trajectory datasets in immersive environment, with a basic interaction tool set. They report that it helps in usability and partially addresses the issue of the steep learning curve of the visualization method.

%Other visualization tools take advantage of immersive technologies for multidimensional data.

%\subsection{Immersive and Interactive Interfaces}  

%A number of 3D data visualization tools have been developed using immersive environments, however, they are tailored to the type of data to visualize. 

%\remove{Specialised in the visualisation of statistical non-spatial datasets, DXR\cite{sicat_dxr:_2019} and IATK\cite{cordeil_iatk:_2019} are visualization tools for the Unity 3D platform. They both use visualisation and interaction grammars inspired by Vega-Lite \cite{satyanarayan2017vega}, to simplify the creation of more or less complex views.  DXR\cite{sicat_dxr:_2019} proposes a high-level GUI, easy to learn and efficient to prototype and personalize quickly visualizations, according to a panel of templates such as 3D bar charts or scatter plots, and more advanced views can be built using DXR's base implementation. Interactions such as scaling, rotating and details on demand when pointing any of the data item visualized are available, but performances decrease after an order of magnitude of a thousand objects. IATK\cite{cordeil_iatk:_2019} proposes similar functionalities, but is more oriented toward scalability, supporting datasets of millions of items and a more expressive grammar for visualization designs.}

This literature review highlighted the lack of methods for the visualization of 3D spatio-temporal data, and especially for surface-mesh based datasets~\cite{kim_comparison_2017}.
The use of the STH, although slightly explored in the literature, showed a high promise in order to produce various meaningful views with reduced dimensionality but with the potential drawback of generating complex and cluttered visualizations. 
However, the benefits of Virtual Reality systems in terms of depth perception and interaction can overcome such limitations. 
McIntire et al.~\cite{mcintire2014stereoscopic} report in their review that stereoscopic displays are better than desktop displays for tasks of identification and classification of objects in cluttered static environments, overall complex spatial manipulation and spatial understanding. 
These are the type of tasks that can be expected during the exploration of complex static visualization, such as STCs, by analysts.

%The extent of the existing works motivated us to create a visualization method for 3D spatial and temporal datasets usable for all type of time-varying 3D spatial data, notably surface-based datasets, taking advantage of the Space-Time Cube concept as well as the benefits of immersive environments.

\section{Space-Time Hypercube Visualization}
\label{STCsection}

This section presents our main visualization method, based on the Space-Time Cube. First, we present a generalization of the STC in 4 dimensions, the STH, then we detail the design choices for the generation and visualization, as well as the interaction methods available. 

As a support example for this section, we will use a dataset of a time-varying 3D spatial data, a 6-hour-long live recording of an embryo development. In this work, we use 100 of the 180 time points recorded, from a 65 to 383 cells embryo. Figure~\ref{fig:teaser} shows some of the time steps that were used.
The dataset is composed of surface meshes, for each cell at each time step. The objects are close together, resulting in a quite dense view. 
The tracking of cells in time, notably after division, are also present, as well as quantitative and qualitative information. 
Movements between time points are of a lower order of magnitude than the size of the cells.
The dataset will be described more thoroughly in Section~\ref{usecases}. 

\subsection{Generalization in 4D} 

\begin{comment}
\begin{figure*}[t]
\centering
\includegraphics[width=\textwidth]{figures/colormapping.png}
\caption{Visualization process. Figure \circled{1} corresponds to the output of the STC generation. Each voxel contains the cell identifier $c_j$ and a time point $t_i$ can be found. Figure \circled{2} is a texture containing quantitative information - here the volume of each cell at each instant - normalized between 0 and 1. The value at $(c_j, t_i)$ points to a 1D color gradient, shown in figure \circled{3}. Figure \circled{4} is the resulting visualization after shading and color mapping. Smaller cells can thus be identified by a colder color, bigger cells by a hotter color.}\label{fig:visualizationFlowDiag}
\end{figure*} 
\end{comment}

%\ferran{Data lies in a 3D or a 4D space, but the "data" visualized can by 1D, 2D ..., e.g. density of the volume, density + cell volume.}
In a classic 3D STC, the data visualized is composed of 2 spatial dimensions as well as a temporal dimension $(x,y,t)$. In contrast, a STH can contain 3 spatial dimensions and a temporal component $(x,y,z,t)$. 
With this abstraction, any elementary operations applicable on a STC, notably described by Bach et al.~\cite{bach_review_2014,bach_descriptive_2017}, should be extendable and applicable to the STH.
Geometry and content transformations now modify a 4D volume, filling operations now includes 4D interpolation, extraction operations can render 4D volumes and flattening operations can compress 4D data into a 3D volume. 
Complex operations for a STH can be designed based on this extended set of operations, in order to provide meaningful visualizations of the data, but with varied points of view and dimensionality, notably including time.
Although, several constraints appear with this additional dimension. 
First, while a STC can be directly rendered, a STH cannot be directly rendered as is. 
To enable a direct visualization, a projection operation on the STH has to be defined in order to yield a 3D visualization.
Second, the design of a 4D operation can be difficult because of the difficulty to mentally visualize a 4D volume.

We tried to design such operation on our example dataset.
%
%A flattening operation as the one used in Trajectory Mapper~\cite{lange2017trajectory} to project 3D trajectories on a 2D plane before extruding on the time axis, could be used to reduce the dimension of the STH by one.
Initially, we considered a flattening operation on the spatial dimension, projecting the 3D snapshots of the dataset into 2D planes, in order to reduce the dimension of the STH by one.
However, because of the topology of our example dataset, using this method on such dense data can result in a volume which would be prone to occlusion issues, because of the increased density due to data aggregation.
In contrast, a volume extraction operator would enable the reduction of the data dimension without the increased data density.
Thus, we opt for a volume extraction operating by defining a hyperplane laying in a 4D space. Nevertheless, two major issues remain: 

\begin{itemize}
    
    \item{An extraction operation implies a loss of data during the creation of the visualization. Thus, the user must have control over which data is selected.}
    
    \item{The concept of a 4D hypercube is not easy to comprehend; the extraction of a volume along a hyperplane is even more difficult. The complexity of the model will be a constraint during the extraction operation, as well as any other interactive operation based on the 4D data.}

\end{itemize}

These two constraints will influence our design choices, especially the generation process for our 3D STC visualization.

\begin{figure*}[t]
\centering
\includegraphics[width=\textwidth]{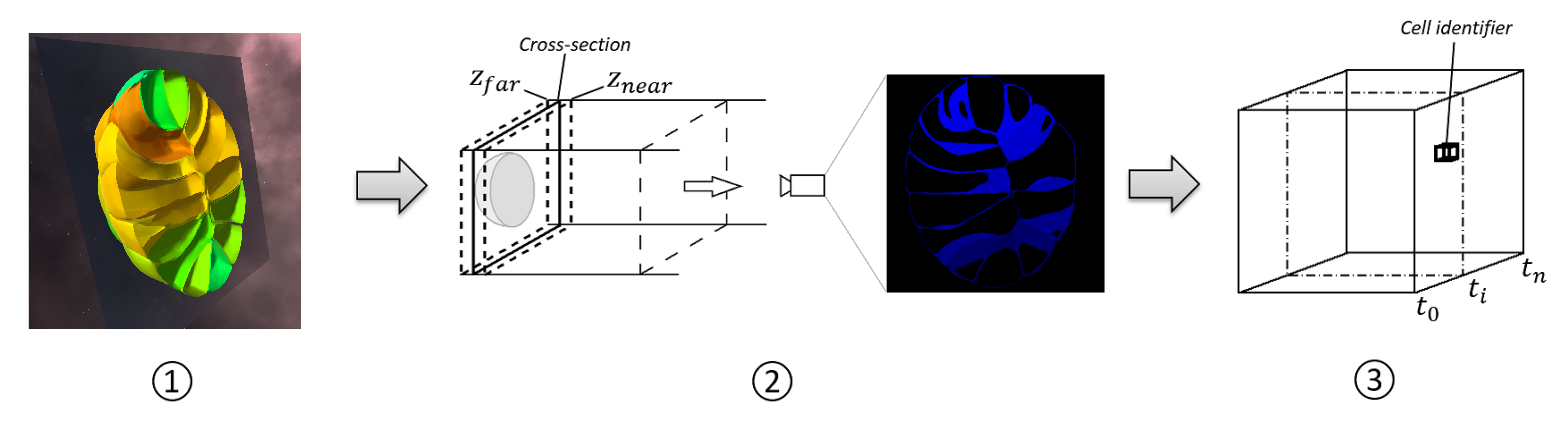}
\caption{Flow diagram of the STC generation. In step \circled{1}, the user places the interactive clipping plane to get the desired cross-section. In step \circled{2}, camera parameters are automatically set in order to render the cross section at each time point. The image presents the output of the rendering operation, using the RGB channels to save cell identifiers. Stacking the rendered images yields a 3d volume, as shown in step \circled{3}. Each voxel contains a cell identifier, and its depth indicates a time point $t_i$. }\label{generationFlowDiag}
\end{figure*}

\subsection{Extracting a 3D STC from a 4D STH}
\label{generation} 

The proposed projection method relies on the use of a 4D hyperplane, manually defined by the user, to extract the 3D visualization.
However, the definition of a hyperplane of this nature is an abstract task which cannot be directly visualized. A simple example would be to define a hyperplane perpendicular to the time axis which would result on the extraction of the spatial data at a given time point. 
In order to ease the process and obtain a hyperplane yielding more informative STCs, we propose a user defined approach based on the cross-section, a commonly used tool in scientific visualization to explore spatial 3D data, extracting a 2D view from 3D data.

%\gwendal{If defining a hyperplane perpendicular to the time axis would straightforwardly yield the classic view of the 3D data at a particular time point, the task becomes way more abstract and complex when the hyperplane actually extends in all four dimensions.}
%\removed{Conceptually, this operation is complex due to the 4 dimensional nature of this type of objects.} 
%
%We propose a simpler approach, based on an extension of the cross-section, which is  

Precisely, the user can place an interactive clipping plane, that will act as a cutting plane, as they would do to get a cut-away view of 3D spatial data. With the same analogy that we used to consider our spatio-temporal data as a 4D volume, the plane, as a time-varying object in the 3D space, can be considered as a hyperplane in the 4D space. 
The operation thus corresponds to a projection such as $(x,y,z,t) \rightarrow (x',y',t)$, with $x', y'$ the projected coordinates on the cutting plane. 
Such projection avoids any spatial distortion that could impair the interpretation of the output shape.
Considering the whole dataset as a 4D Space-Time Hypercube, it sums up as a space cutting~\cite{bach_review_2014} by a hyperplane, extracting a 3D space-time cube.

%We use a capture method based on rasterizing the 3D content. The main benefit of such method is that it is independent from the type of 3D spatial information mapped, would it be volumetric data or surface meshes.

The proposed generation method is data agnostic (mesh-based or volumetric) and its three main steps are illustrated in Figure~\ref{generationFlowDiag}.
\circled{1}~The user places a clipping plane on the model at any time point in order to define the ``hyperplane'' which will determine the projection. 
\circled{2}~For each time step, a 2D image of the cross-section of the 3D model is computed. The rendering is achieved by setting an orthogonal camera perpendicular to the clipping plane and setting the near and far planes at a distance of $\epsilon= \sfrac{(z_{near} - z_{far})}{2}$ towards the clipping plane. The field of view of the virtual camera is minimized according to the maximum size of the cross-section over time. 
The color channel of the rendered image encodes objects (cells in our examples) identifiers, which allows indexing numerical or categorical data during visualization.
This requires object information, which implies potentially complex segmentation processing in the case of biology imaging.
If this type of information is not available, raw position data can be encoded in the volume either way.
\circled{3}~The output images are stacked into a 3D texture in which the RGB channel is used to keep track of the object identifier. The depth coordinate for each pixel encodes the time step.
The use of object identifiers is considered for convenience when the spatio-temporal data has this information. In case that this information is not available, other information could be stored, such as density values from a CT or fMRI volumetric dataset.

% \ferran{I know that it was me that asked for this algo... but I'm not convinced now... ^^.}
% \begin{algorithm}[b!]
% \caption{Generation of the Space-Time Cube. First, the camera parameters are initialized and the identifiers of the cells are encoded into colors. Then, for each time point, the camera renders the cross-section and the result is added into the 3D texture.}\label{alg:generation}
% \begin{algorithmic}[1]
% \Procedure{GenerateSTC}{$ray,start,end$} : Texture3D

% \State Texture3D texture3D
% \State InitCameraParameters()
% \ForEach{cell $\in$ Cells}
% \State cell.color $\gets$ encodeIdAsColor()
% \EndFor

% \ForEach{$t \in [\![t_0, t_n]\!]$}
% \State tempTexture2D $\gets$ camera.Render()
% \State texture3D.AddTexture2DAt(tempTexture2D, t)
% \EndFor
% \State \textbf{return} texture3D

% \EndProcedure
% \end{algorithmic}
% \end{algorithm}

\subsection{Visualization: 3D rendering and quantitative data}\label{visualizingSTC}

%\ferran{I would orient this part a bit differently, here you can start stating that the data generated by the projection operation is a 3D volume. Then state that a number of works have already focused on the rendering of such data, stating the choices that you did with this respect.}
%
%//TAG : reduce this, too technical
The projection operator generates a 3D texture which can be directly rendered using Direct Volume Rendering methods~\cite{stegmaier2005simple}. 
Due to the nature of our datasets, we consider an opaque rendering avoiding semi-transparency, in order to limit color distortion when using color to encode information. 
We encode numerical or categorical information that has to be rendered in textures that can be pre-computed.
This way, using the object identifier, such information can be mapped directly on the volumetric visualization, and can also be switched at low cost.
We apply a Blinn-Phong illumination model on the cube to enhance depth cues and get a better understanding of the shape of the volumetric data. 
To reduce the rendering time, normals are pre-computed during the generation of the STC and stored in an additional 3D texture. Normals are computed using 3D Sobel-Feldman gradient operators on each direction.
The rendered STCs, before and after color mapping, are shown in the second and third images of Figure \ref{fig:teaser}.

However, the STC could still contain a lot of information and could present a complex layout. The following section details manipulation techniques proposed in order to improve the exploration of the STC.

%\begin{figure*}[t]
%\includegraphics[width=1\textwidth]{figures/VR_framework2.png}
%\caption{Examples of tool uses: the first one allows the creation of a scatter view \circled{1}, with green lines representing adjacency links. The cube \circled{2} is a two-handed control allowing rotation and rescaling of the embryo. The clipping plan \circled{3} is remotely controlled by the user’s hand. The laser pointer \circled{4} has the list of quantitative and qualitative information displayable, here switching from Volume to Remaining Lifespan (i.e. before its next division) of the cell.}\label{fig:VRframework}
%\end{figure*} 

\begin{figure*}[t]
\begin{center}
\includegraphics[width=0.85\textwidth]{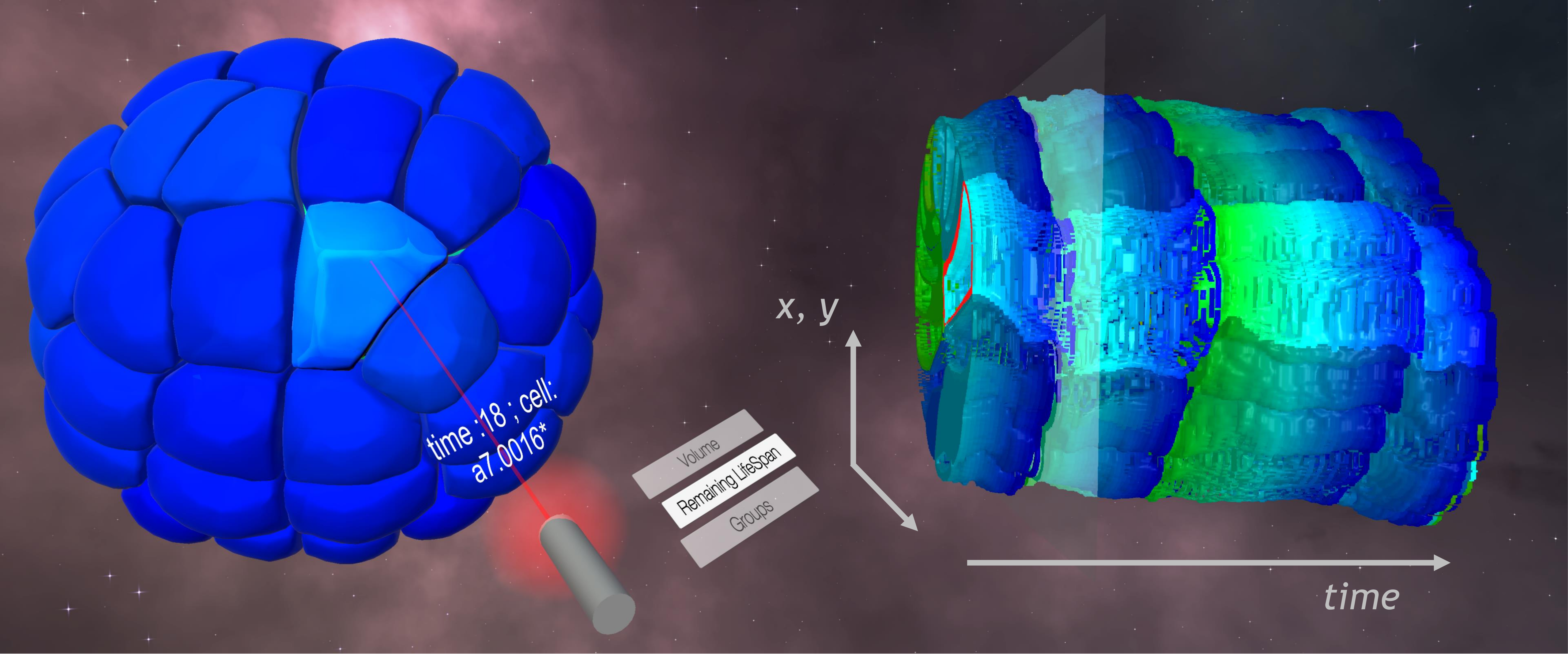}
\caption{STC (right) and meshed model (left) visualizations in the VR framework. On the laser selector, a list of available information allowed the user to color map the remaining lifespan of cells on the visualizations. A cell is selected on the meshed model. The name of the cell is displayed on the pointer, and feedback appears on the STC, highlighting the lineage of the cell.}\label{fig:changeInfo}
\end{center}
\end{figure*}

\subsection{Interaction} \label{STCinteraction}

Interaction will be a key concept to compensate for issues implied by direct rendering, such as occlusion. Bach et al.\cite{bach_review_2014,bach_descriptive_2017} described several generic operations that could be applied to a STC, and mentioned dynamic uses of those operations.  
In addition, by displaying the STC in a VR environment, additional depth cues will be available, such as binocular cues and motion parallax cues~\cite{loomis2003visual}, which will increase the identification of the intricate structures in the STC.

In the first place, as previously mentioned, the STC is interactively generated, since the user can choose the base cross-section, and thus the hyperplane, and start the generation process at run time, doing a space-cutting operation of the 4D data.
Taking advantage of immersive environment in terms of interaction, we designed a number of interactions adapted to our data structure and the benefits it would imply for data visualization.
All the techniques described in the following sections assume that the user is wearing a head mounted display (HMD) and holding by the hands two 6DoF hand-held controllers. As such, the user will be able to explore the STC by just naturally walking in the virtual environment and use both hands to use the different tools.
We refer the reader to the accompanying video which showcases the different interaction tools.

\subsubsection{Exploration tools} \label{explorationTools}

The tools described in this section are meant to help the user explore potentially dense and occluded volume representations.

\textit{Light: }The user can grab and move a point light source, represented as a yellow sphere.
This helps understand the shape of the volumetric data by adding depth cues, reveal shape, reliefs and details of the STC.

\textit{Clipping Plane: }An interactive clipping plane generate cut-away views of the STC.
The user can directly grab the clipping plane and manipulate it using remote hand translations and rotations. Some orientations are easier to interpret: 
    \begin{itemize}
        \item{Time cutting, i.e. putting the plane perpendicularly to the Z-axis, corresponds to going through the original cross-section at given instants.}
        \item{Linear space cutting, i.e. putting the plane along the Z-axis, corresponds to the evolution of a segment of the spatial data over time, giving for instance information of the evolution of the spatial structure.}
        \item{In case of moving objects, slightly oblique cutting can help track the object over time. This type of cut was used to generate the cut in Figure \ref{fig:deformation}.}
    \end{itemize} 

\textit{Selection operator: }The user can point and display summarized information of a hovered object, such as its identifier. 
This is a detail-on-demand operator that can select and display additional information on the object on click.
%
%It appears as a red arrow object, so that the user can put the pointer precisely on the desired object, providing a highlight feedback upon hovering or selecting.

\textit{Switching the color mapped information: }When various numerical or categorical information are present in the dataset, the user can scroll through the list of information displayable (cf Figure \ref{fig:changeInfo}).
As detailed in section \ref{visualizingSTC}, the underlying operation corresponds to switching the texture containing the information. This is thus a light operation allowing interactive use.

\subsubsection{Filtering Operations} \label{filteringOp}

As occlusion remains an issue in dense volumetric data, reducing the amount of data displayed becomes necessary.

\textit{Value Filter: }This tool allows the user to filter the range of continuous data displayed on the STC by using a slider.
Filtering on value can help identify groups of objects sharing same characteristics, complementing the color mapping by removing the occluding other objects.

\textit{Time Filter: }This filtering operation lets the user define a temporal window to constraint the data rendered in the STC within two time steps, selected using a slider.
Events ongoing during this specific temporal window can then be more easily pinpointed.

\textit{Object Filter and Tracking: }Focusing on objects, two issues came up. First, we need to keep track of objects in the temporal dimension. Second, the presence of numerous objects can occlude the visualization. 
The user can solve these issues respectively by highlighting or hiding the objects. It can be done either by pointing the object using the same tool that gives details on demand, described in \ref{explorationTools}, or by selecting an object in a list of identifiers. The selected object thus changes state \textemdash from normal to highlighted to masked \textemdash upon clicking the trigger of the controller.
If the objects follow a tree hierarchy in time, such as the cells with their children cells in our example dataset, the operation is applied to the children objects. Figure~\ref{fig:changeInfo} shows an example of object tracking.
It should also be noted that these options can be used only if object and tracking information are present in the dataset. 

\subsubsection{Linking Multiple Visualizations}

We also took advantage of the large workspace offered by virtual environments to display the meshed model in addition to the STC. 
We followed advice from Munzner~\cite{munzner2014visualization} for the design choices implied by juxtaposition: multiple views should be coordinated, and even interactively coordinated.
Consequently, the operations we described above are applied to the STH, and feedback of these operations is given through the two visualizations displayed - the STC and the meshed representation.
Links can be set up between the visualization by applying a shared encoding of the effect of the operations.

\textit{Linking the time exploration: }In addition to the time filtering sliders evoked previously, a middle cursor controls the current time displayed on the meshed model. Feedback of this cursor appears as a transparent plane orthogonal to the temporal axis of the STC, as shown in Figure \ref{fig:changeInfo}.

\textit{Linking the spatial exploration tools: }
The same color mapping is applied to the STC and meshed model, notably after switching the color mapped information interactively, as shown in Figure \ref{fig:changeInfo}.
The clipping plane used on the meshed model to generate the STC can go back on click to a default position, which is the one of the displayed STC's base cross-section. 
A marker line on this cross-section gives context information about the position of the clipping plane applied to the STC. This line corresponds to the intersection between the STC's clipping plane and the temporal feedback plane.

\textit{Linking the filters: }Any filtering operation done on the STC is also applied to the meshed model, and vice versa. The same objects are highlighted, hidden or filtered out by time or value. 

These operations were designed to help users to explore temporal evolution with the static visualization offered by the STC, and correlate the multidimensional data or pinpoint events by combining the two visualizations and the interactive tools available, as illustrated with the following use case.

\subsection{Performance}\label{performances}

The VR application runs on Unity 2018.2.21 and is supported by a PC with Windows 10, an Intel Xeon W-2104 CPU (4 cores, 3.2 Ghz base frequency) and a RTX 2080 GPU. All interaction methods described previously are designed in order to maintain a framerate above 45 fps.

In terms of performance and texture resolution, the STC used in the figures has a resolution of 256x256x100. The full generation process takes about 5 seconds.
This process is GPU-friendly, which enables the generation of the STC during runtime. Normals are for instance generated and saved instantly using compute shaders. However, the bottleneck of the current implementation of the generation algorithm is the CPU - GPU transfer, provoking a framerate drop. 
    
\section{Use Case: Application for Morphogenesis}
\label{usecases} 

This section presents an application of the STH to the visualization and analysis of spatio-temporal data from the recordings of embryonic developments of a tunicate, a marine invertebrate.
We present the studied dataset, which has been already shown in the previous section, and detail a VR application in which the STH has been integrated. Finally, we describe two use cases illustrating the potential benefits of the STH. 

\subsection{Presentation of the Dataset} 

The data set used to illustrate the STH data is a live recording of the embryonic development of the \textit{Phallusia mammillata}, a marine invertebrate animal of the ascidian class. 

The embryo was imaged every two minutes for six hours using confocal multiview light-sheet microscopy, generating a 4D dataset with isotropic spatial resolution of several terabytes~\cite{guignard2017contact}. 
The data was then segmented using the ASTEC pipeline (Adaptive Segmentation
and Tracking of Embryonic Cells)~\cite{guignard2017contact} and meshed using the VTK library~\cite{schroeder2004visualization} and MeshLab~\cite{cignoni2008meshlab}, producing a surface-based spatio-temporal dataset of a few hundred megabytes representing the embryo. 
Examples of the surface meshed model of some of the 180 time points recorded are shown in Figure \ref{fig:teaser}. 

In addition, the ASTEC pipeline extracted global lineage trees of the embryo, i.e. the tracing of cellular genealogy derived from cell divisions and migration.
Other categorical or continuous data, such as the volume of cells or the remaining lifespan - i.e time before the next division - were computed or added by the community of experts working on the dataset. 
We also had access to the ANISEED\cite{brozovic2017aniseed} database, providing us with categorical information about the expression of various genes in the cells.

\subsection{VR Visualization Application} \label{framework}

Taking advantage of the large workspace offered by immersive environments, we developed a VR framework for the user to interact with the meshed model, in addition to the STC visualization. We based our environment on the framework \textit{MorphoNet}~\cite{leggio2019morphonet}, an online interactive browser for the exploration of morphological data. 
The development of the application was done using Unity 3D and the HTC Vive was the main visualization and interaction system.

Our framework displays the dataset described above using meshed model of the embryo at each time point, as well as a STC based on a cross-section placed close to the median plan of the embryo, as shown in Figure \ref{fig:usecasemodel}. 
The color mapped on the visualizations here corresponds to the volume of the cells.

Regarding the user interface, a virtual desk contains all the available tools that the user could grab using the trigger button of the HTC Vive controller.
Once a tool is grabbed, it can be used either using the trigger, or by performing circular motions on the circular touch pad of the HTC Vive controller. This last interaction allows to control a continuous value (e.g. the time) or to scroll on a list.
The user could grab a tool in each hand.

\begin{comment}
\begin{description} 
    \item[Laser Selector: ]{A laser pointer allowed the selection of a cell using raycasting selection. 
    %
    Similarly to the tool described in \ref{explorationTools}, this pointer displays summarized information of the hovered object, and can highlight or hide it.
    %
    Note that the list of information is also displayed on this tool.}
    
    \item[Cell Information Display: ]{Information regarding the selected cell appears on a virtual screen placed on top of the desk. }
    
    \item[Clipping plane: ]{A semi-transparent plane manipulated through hand motions does a cross-section of the model. The user can interact with the plane remotely by pressing the trigger button. The STC generation process is based on such cross-section.}
    
    \item[Scatter view: ]{Creates a scattered view of the cells of the embryo. Green lines will appear after a threshold to keep cues of adjacency between the cells. The scattering parameter can be changed by doing circular motions on the touch pad.}
    
    \item[STC generation: ]{As described in section \ref{generation}, the STC visualization can be generated during runtime, according to the placement of the clipping plane tool presented above.}
    
\end{description}
\end{comment}

The different designed interactions were a \textbf{laser selector} to select a cell (see Figure \ref{fig:changeInfo}), a screen to \textbf{display information} on the selected cell and a remote control \textbf{clipping plane}, also used to select the plan for the STC generation.
In addition, a bi-manual hand manipulation, similar to a remote Handlebar technique \cite{song2012handle}, allows \textbf{applying rotations} to the meshed model by holding of the trigger button on both controllers. This can be enabled at any time, even while using a tool. 
The meshed model is virtually attached to an axis defined by the position of both hands. In order to activate the rotation, users have to press the trigger of both controllers.
The modification of the length of the user’s hands gap modifies the scale of the meshed model.
The rotation can also be applied to the STC, depending on which visualization the users is looking at.

This set of tools, as well as the ones dedicated to the STC, provides the user with primary controls to explore the dataset.

\subsection{Use Cases} \label{xpUseCases}

To study functional organization and arrangement of the tissue, embryo developmental studies remain essential to access cellular functions inside a self-organized isolated system.

With a fast development, a few hundreds of cells, ascidian embryo is the perfect choice to analyze the link between cells or tissue geometries and differentiation. In the case of \textit{Phallusia mammillata}, the cell membranes are very transparent, which makes for an easier imaging of the membranes or endosomes of the cell through light sheet microscopy.
Without any apoptosis, i.e. programmed cell death, or cell migration in early ascidian development, embryo topological complexity can be summarized in the result of unequal divisions of cells, resulting in two different volumes for each daughter cell, and/or asynchronous divisions, i.e. two daughter cells having different cell cycle duration. Each of these events correlates with cell fate decision of the cells.
This way, with the exploration of cell architecture and adjacency in a dynamical view in embryos, we have access on a major part of the story and decisions taken by each cell.

We will illustrate in the following examples of use of the STC to observe events and behaviors with a strong temporal component related to the exploration issues described above.
For this purpose, we generated a STC from the embryo cross-section shown in Figure \ref{fig:usecasemodel}.

%\begin{figure}[t!]
%\centering
%\includegraphics[width=0.45\textwidth]{figures/cellViz.png}
%\caption{This visualization shows the cell at all time points. The cell deformation over time can be estimated by direct comparison between the occurrences of the cell, and a cell division is also highlighted.}\label{fig:cellViz}
%\end{figure} 

\subsubsection{Cellular Deformation} \label{UCdeformation}

One of the main steps of the embryonic development for most animals is gastrulation. It is characterized by morphogenetic movements that form a cavity in the embryo. 
These movements set the base of the determination of the morphology of the future individual. The actual source of these movements and deformations of the cells is still unsure, and no formal measure of deformation exists. 
Visualization of this deformation over time is necessary for the analysis of the embryonic development. 

\begin{figure}[t!]
\centering
\includegraphics[width=0.8\textwidth]{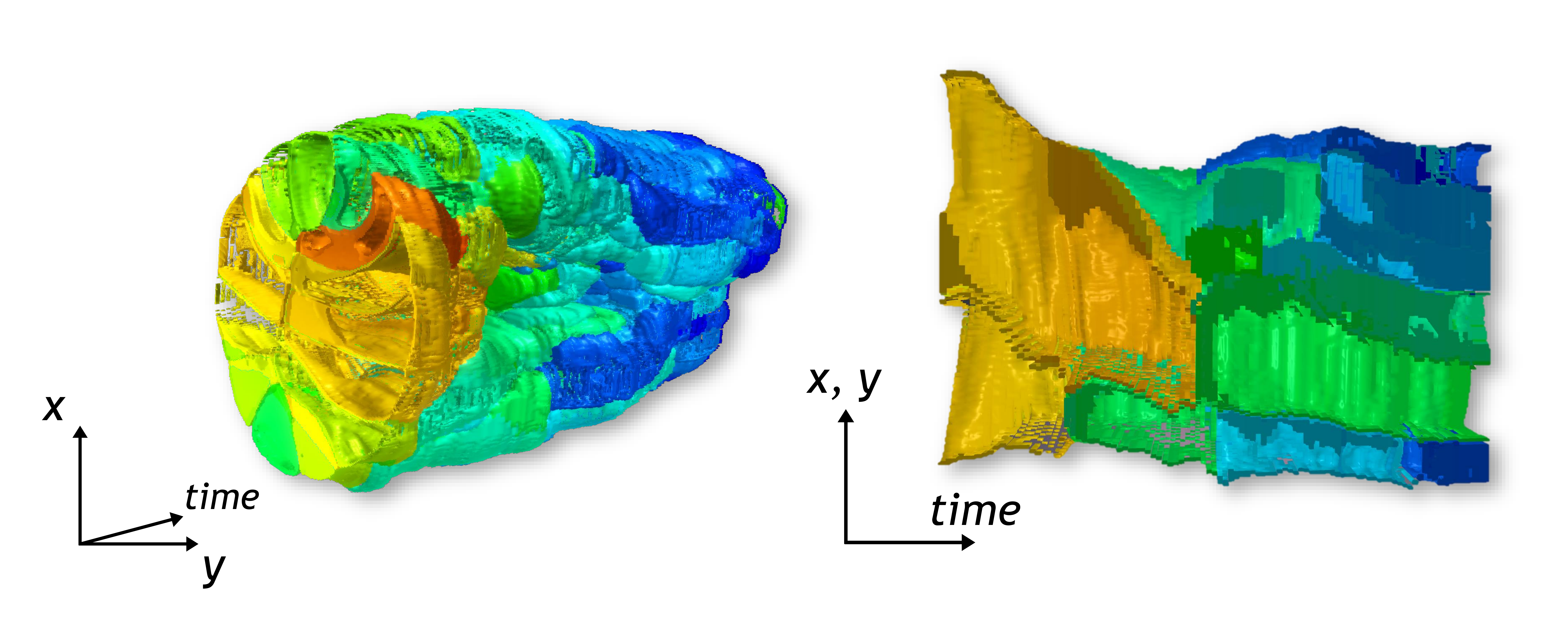}
\caption{The STC at the left presents the cavity formed by gastrulation. The cross-section at the right shows the deformation of cells involved in this particular morphogenetic movement.}\label{fig:deformation}
\end{figure}

The STC shows the deformation of the embryo over time, as shown in Figure \ref{fig:deformation}. A cross-section allows the exploration of the main cells involved in the movements and provides a view of the global and local deformation. 
The user can then identify the cell to visualize by selecting them on the STC and continue the exploration on the 3D representation.

\subsubsection{Waves of Cell Divisions} \label{UCwave}

During the early stages of the embryonic development, cells tend to divide quite simultaneously. This phenomenon results in ``waves'' of division, during which numerous cells divides within a short time. 

By sliding through the time component, one can notice this kind of event on the surface-based representation through several aspects. 
First, the characteristic dynamics of the membranes of a cell can be observed on a large number of cells: on the surface of the embryo, and inside by using a clipping plane. 
Second, as cells divide, they reduce their volume. Thus, the color mapped on the embryo will change a lot during such event.

\begin{figure}[t!]
\centering
\begin{minipage}{.45\textwidth}
  %\centering
  \hspace{-0.1\textwidth}
\includegraphics[width=1.2\textwidth]{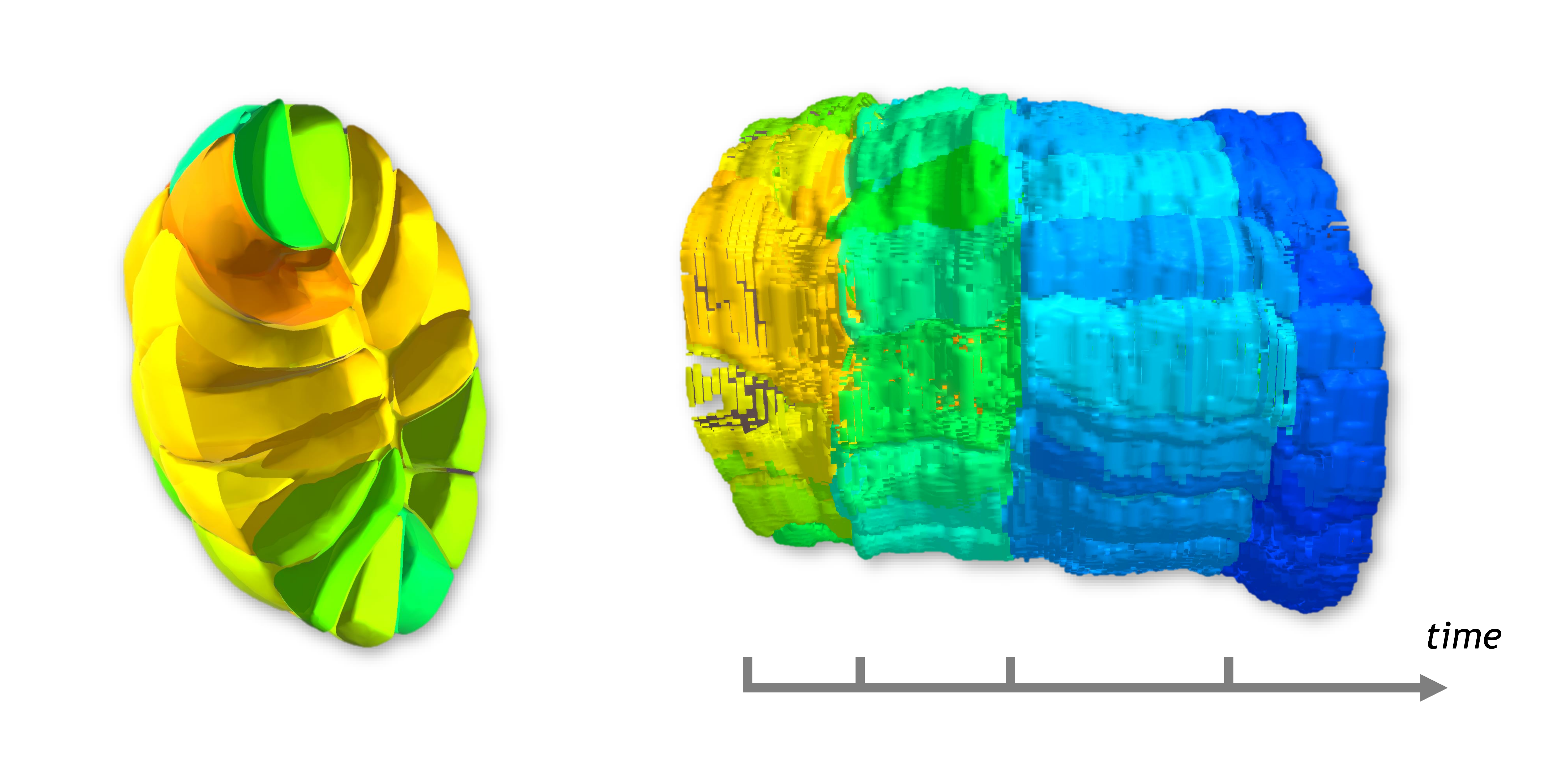}
\caption{Cross-section and related STC generated. The color displayed on the STC corresponds to the volume of cells, and helps distinguish 4 parts.}\label{fig:usecasemodel}
\end{minipage}
\hspace{0.05\textwidth}
\begin{minipage}{.45\textwidth}
  \centering
\includegraphics[width=0.94\textwidth]{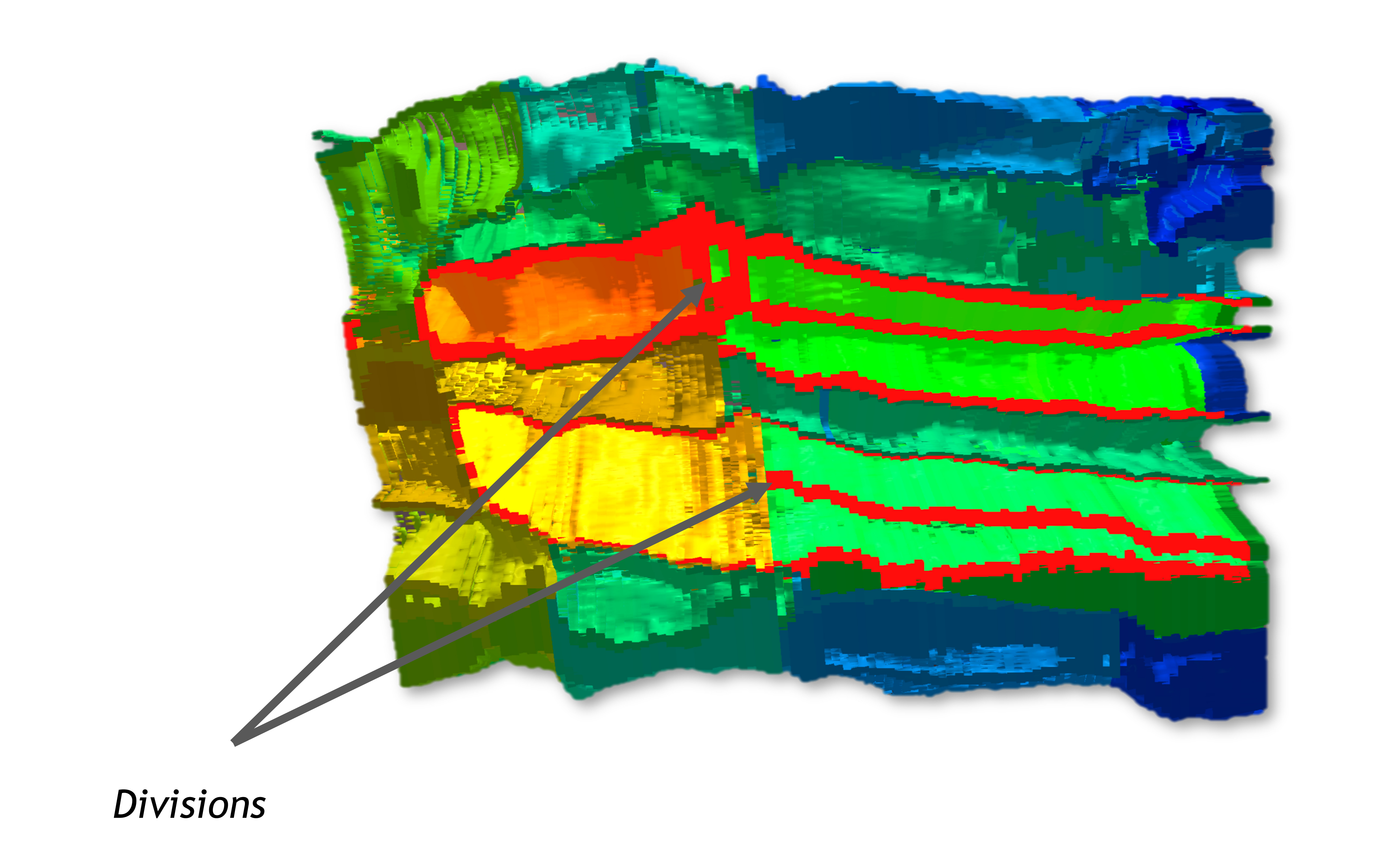}
\caption{Cross-section of the STC. Two divisions are indicated. The membrane of the original cells and their respective daughters cells are highlighted in red.}\label{fig:asynchrony}
\end{minipage}
\end{figure}

Keeping this last property in mind while exploring the STC, color clusters become distinguishable thanks to the volume information displayed. 
This observation can thus be done directly, without having to do a time-sliding operation, as shown in Figure~\ref{fig:usecasemodel}.
Up to four color clusters, hence three waves of division, can be identified.

\subsubsection{Asynchronous Behavior in Cell Divisions}\label{UCasync}

Regarding cell determination, cells with different fates will not result in the same number of cells, and embryonic development has to be controlled. 
Thus, at specific moments, some cells will not divide at the same time as the other ones and skip a wave of division. 

Exploring the surface-based representation in order to find cells presenting this behavior, the user has to slide back and forth in the time window corresponding to the wave of division and observe which cell will not divide.

The STC can give a more global view of this time window, with a part of the context of the embryo. 
In Figure~\ref{fig:asynchrony}, membranes of a cell are highlighted in red. 
Only one division can be observed for each of these cells during the time window represented.
Up to three were to be expected here, according to the waves of division identified before, which confirms an asynchronous behaviour for these cells.
With the feedback of the cell selection tools on the STC, the user can find quickly the related cell on the 3D model of the embryo.

\section{User Evaluation} \label{evaluation}

In this section, we present and discuss a user evaluation aiming to assess the benefits and limitations of the proposed method using the previously defined use cases.
The evaluation was motivated due to the complex and intricate 3D visualizations that are generated by the proposed projection, which could be difficult to interpret.
Thus, the main goal of the evaluation was to provide a first assessment on whether and how users could take advantage of the generated STC visualizations when punctual and continuous events have to be located.
Although other use cases could be envisioned, we decided to focus on embryology use cases, as they were our first motivation for proposing this approach for the projection of the STH. 
In this context, as biologists mainly explore this datasets by exploring one time point at a time, as a baseline condition, we considered the mesh visualization on which users can navigate in time (e.g. using a slider); hereinafter referred to as the meshed model alone condition.
In the meshed model \& STC condition, the mesh data was also juxtaposed to the STC visualization.
%Users cannot really know how to set the right clipping plan from their first use of the method, which could introduce bias. -> steep learning curve on STCs, STH voilà quoi. For the evaluation, we generated one STC in order to get similar amount of information from the visualization for each user -> one objective criterion, we took the deformation showed in the figures. The evaluation is about interacting with dimensionally unusual visualization, in juxtaposition with a usual 3D representation.

\subsection{Tasks and Hypotheses}

We focused our evaluation on the efficiency and accuracy provided by the STC through tasks of identification of cell divisions, waves of divisions and asynchronous behavior, as described in \ref{xpUseCases}. 
Such events and behaviors have temporal and spatial components, thus can be considered as 3D temporal objects. 
Using the same abstraction as described in section \ref{STCsection}, they qualify as 4D objects. 
As such, we suggest that one of the visualizations will render more information depending on the characteristics of the event. 
We discern two type of events. First, punctual events can be identified by comparing the state between 2 time points. To detect such events, the user requires more complete spatial contextual information over a very local temporal context, i.e. a few time points. For this reason, we hypothesize that the meshed model visualization will be more helpful is the detection of punctual events.
The second type of events evolves over time, and is thus mostly characterized by a temporal component extended in numerous time points. As the generated STC provides complete temporal information over a reduced spatial context, we hypothesize that the use of the STC will be more relevant to detect such events.

For this study, we considered the complexity of the tasks and the fact that our participants were not domain experts. Consequently, we chose to focus our evaluation on the accuracy of the analysis and not on efficiency, which could be more biased, notably by the approach taken by the participant.
Based on our previous rationale, our main hypotheses were: 

\begin{description}
    %\item[H1:]{Punctual events in time will be pointed on the meshed model more accurately than on the STC;}
    \item[H1:]{Punctual events in time will not be pointed more accurately using the STC than on the meshed model only;}
    \item[H2:]{Events or behaviors with a larger temporal component will be identified more accurately using the STC than on the meshed model only.}
\end{description}

\subsection{Apparatus and Participants}

The PC configuration is the same as described in \ref{performances}. The virtual environment was displayed in a HTC Vive Pro HMD, with a resolution of 1440 x 1600 per eye. Two Vive Pro controllers were available, including 3 buttons and a touchpad. The HMD and the controllers were tracked in a space of a 2.5m x 2.5m surface.

12 participants, 10 men and 2 women, were recruited from the local laboratory for this experiment, aged from 22 to 32. They all had medium to high experience in VR environments, but had no expertise on the dataset nor in embryology.

\subsection{Experimental Task} \label{taskDesign}

In order to keep the morphogenetic analysis context, we designed tasks of annotation accessible to non-expert users.
From the use cases presented in section \ref{xpUseCases}, tasks of pinpointing waves of cell divisions or identifying asynchronous behavior in cell divisions corresponded to events with large temporal component. 

As a task of pinpointing punctual events, we considered cell divisions. 
They are common events yet essential for the analysis of the embryonic development.
The temporal resolution of the dataset allows us to consider the divisions as punctual events. 
The spatial deformation occasioned by a division can be easily noticed on the 3D representation of the embryo by sliding through time.
An additional hint would be the abrupt change in the color displayed, which can be used to pinpoint this event on the STC. 
As described in \ref{filteringOp}, the hierarchy, corresponding here to the cell lineage, is highlighted, which can also help identifying a cell division on this visualization.

\subsection{Experimental Protocol}

At the beginning of the experiment, participants signed an information consent form and were briefed regarding the equipment used, the data recorded and the experimental tasks. 
In order for the participant to understand the context of the tasks, we explained the nature of the dataset, specifying the methods of recording and the purpose of analyzing such data. 
Emphasizing on the temporal component of the data, we then presented the STC visualization, its generation process and its purpose in terms of temporal visualization. 
Since the participants were non-experts, we explained for each task the characteristics of the events to pinpoint, the scientific purpose in embryology analysis as well a few methods to identify those events in each visualization. 
Finally, we presented the virtual environment and the available tools to the participant for about 8 minutes, before having themselves equip the HMD and get familiar with the framework, for about 8 minutes as well.  Participants were then asked to perform the three exploratory tasks: 
\begin{itemize}

\item{Discern the number of waves of cell divisions, providing an approximate estimate of the time for each wave of divisions; duration of the task: 2 minutes}
\item{Find cell divisions and give corresponding instant and identifier; duration of the task: 4 minutes}
\item{Find cells with asynchronous division behavior and give corresponding identifier; duration of the task: 4 minutes}

\end{itemize}
For the tasks of identifying divisions and asynchronous behaviors, the events were present abundantly enough not for anyone to finish the task early.
The experiment was divided in two blocks. Depending on the visualization method available, the participant performed the tasks first with the meshed model alone and then with meshed model \& STC, or vice-versa. 
The order of the visualization method was counter-balanced to minimize ordering effects. 

The main goal of the experiment is to evaluate how users will take advantage of the generated STC visualizations.
We thus decided to provide an initial STC visualization.
In addition, the participants worked on the same dataset and the same STC view, generating smaller variability and enabling better comparison of results between participants.
Second, considering the potential learning curve in STC based visualizations, we assumed that the level of comprehension of the STH would be very different from user to user.
Such differences could induce an important bias in the experiment, especially since the users are not familiar with the datasets observed.
We thus generated a STC arbitrarily, though following one objective criterion: the plane would be placed in order to capture the gastrulation movement described in part \ref{xpUseCases}.
We expected this to act as a point of reference to help participants apprehend the data.
Nonetheless, we left the opportunity for the participants to generate another STC from a different cross-section, yet none of them did.

For each condition, we recorded the answers of the participant for each task, the total time use of each tool, and the total time watching either visualization.
At the end of the experiment, participants had to fill a short questionnaire in order to gather subjective impressions of the overall app and demographic data (age, gender, and VR experience).

\begin{figure}[t!]
\centering
\includegraphics[width=0.8\textwidth]{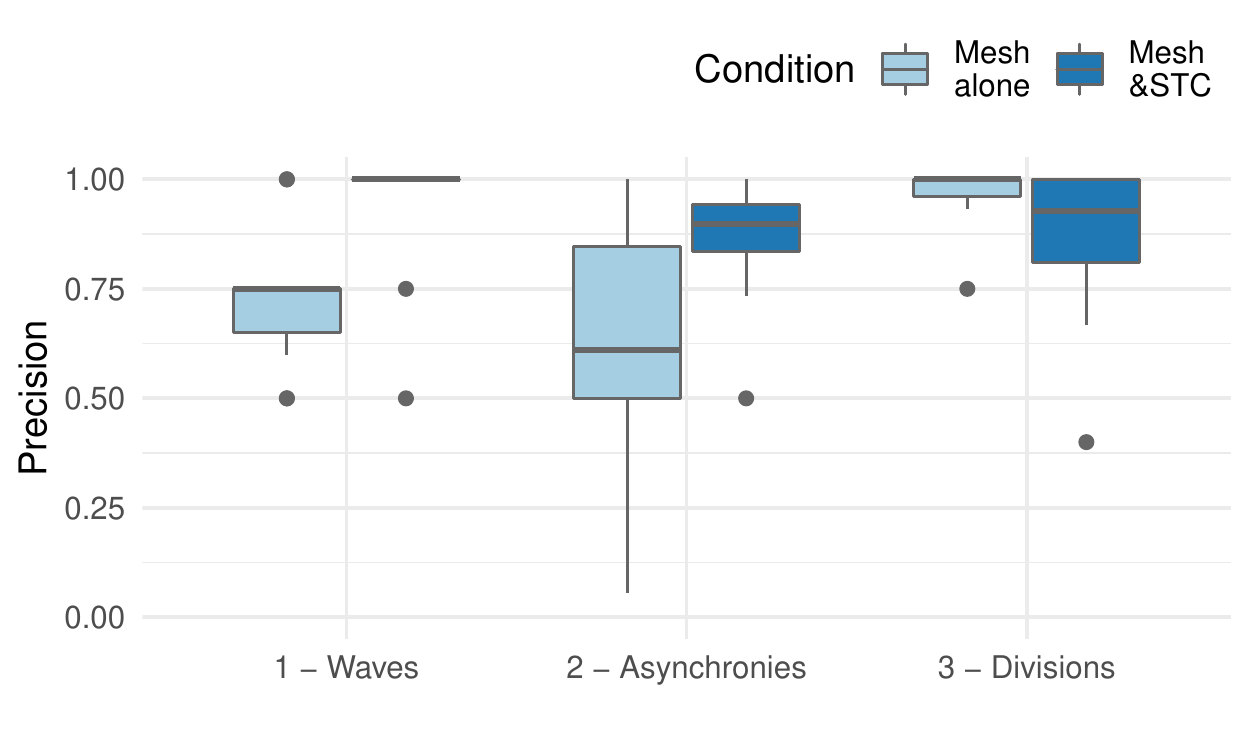}
\caption{Box plots of the precision of the participant for each tasks and condition, i.e. with or without STC. Each plot corresponds to 12 observations.}\label{fig:boxplotPrecision}
\end{figure}

\subsection {Results}

Eventually, we get a total of 24 paired observations, characterized by 2 two-level factors, the \textbf{order} of the observation and the \textbf{condition} - i.e. whether or not the STC is present.
Focusing the study on accuracy, we computed the precision (true positives on total selected elements) for each task.
Repeated measures two-way ANOVA was used to analyze the results, considering the order as a between-subjects factor and the technique as a within-subjects factor. When the normality assumption was not met (Shapiro–Wilk normality test), the Aligned Rank Transform (ART) was used before performing the ANOVA analysis.
%When necessary, Tukey HSD post-hoc tests were performed.
%
The statistical analysis was done using R language.

%boxplot with each precision tests

%emmeans -> interpret the result
%two-way anova / wilcoxon pairwise to talk about the test result

%structure : what is tested / test to exclude effects / test to validate the hypothesis / interpretation

\paragraph{Waves of Cell Division Counting Task}
The two-way ART ANOVA order and condition vs precision showed a significant effect of the condition (\anovatest{1,10}{24.72}{0.001}{0.64}), and no significant order nor interaction effect. 
Post-hoc tests showed a significantly ($p<0.001$) better performance in terms of accuracy using meshed model \& STC (\msd{0.94}{0.16}) rather than the meshed model alone (\msd{0.72}{0.16}).
This \textbf{supports H2}.

\paragraph{Asynchronous Behavior Identification Task}

The two-way ANOVA order and condition vs precision showed a significant effect of the condition (\anovatest{1,10}{5.61}{0.05}{0.36}), and no significant order nor interaction effect. 
Post-hoc tests showed a significantly ($p<0.05$) better performance in terms of accuracy using meshed model \& STC (\msd{0.87}{0.14}) rather than the meshed model alone (\msd{0.64}{0.27}), \textbf{supporting H2}.

\paragraph{Cell Division Identification Task}
The two-way ANOVA order and condition vs precision showed no significant result on any factor nor interaction, so we can give \textbf{no conclusion about H1.}
The box plot of the precision of the participants on this task (cf Figure \ref{fig:boxplotPrecision}) does not show important difference either.

\subsection{Discussion}
The results from the \textit{waves of divisions} and \textit{asynchronous behaviors} tasks, considered as objects with large temporal component as explained in \ref{taskDesign}, showed that the use of the STC provided higher accuracy in those identification tasks, \textbf{supporting H2}.
However, these tasks were designed with numerous known targets but unknown location in space and time.
Placing the user in more diverse situations would help to estimate more thoroughly the added value of the STC in general exploration tasks.
The results from the cell division task gave no significant results, allowing \textbf{no conclusion on hypothesis H1}. 

As evoked by our hypotheses, we focused this study on accuracy. For various reasons, the protocol as it is could not allow a significant efficiency evaluation.
Participants were given the same advice on how to approach the tasks, described in \ref{xpUseCases} and \ref{taskDesign}. However, the numerous tools available made it difficult for them to choose an optimized strategy.
Users usually chose a strategy at the beginning of a task and followed it until the end, even if it felt suboptimal.
Such phenomenon creates subgroups. To counter this effect, we would require either more participants, or ones that are trained with the system.
Furthermore, we focused our study on only one dataset, since it was the most complete one at our disposal. 
Considering our task design, we could expect an important order effect for our efficiency measurements, since users would remember what they did on the same task in the other condition.

Despite all these constraints, we could output a result on the efficiency of the system. 
On the \textit{waves of division task}, some users could finish at least 20 seconds before the 2 minutes time limit. 
In the meshed model only condition, only 1 out of 12 finished the task early (0 on the first part, 1 on the second part). In the meshed model \& STC condition, 8 out of 12 finished the task early (4 on the first part, 4 on the second part).
The two-way ANOVA order and condition vs the boolean value "finished the task early" showed a significant effect of the condition (\anovatest{1,10}{14.41}{0.01}{0.59}), and no significant order nor interaction effect.

\section{General Discussion}\label{discuss} 
%%Quick look-back on the experiment

%The evaluation of our STC visualization seemed to support that events with large temporal component can be identified more accurately.

While the evaluation described in the paper have showed the potential usages and advantages of the STH for the visualization of 3D spatio-temporal data, there is yet a number of additional future works in order to cope with its current limitations. 
%
%%STC flaw
First, the generated STC visualizations are based on a projection operator which enables the visualization of a subset of the original 4D data. Although this approach generates a compact representation of the spatio-temporal data, it also results with a loss of spatial information and can be sensible to strong motions within the dataset. 
Strong motions can be particularly problematic for objects close to the cutting plane that can result in wrong interpretations (e.g. an object shrinking when it is just moving away from the clipping plane).
In the current implementation of the system, we tried to compensate for this negative effects by the juxtaposition of the meshed model, which gave additional contextual information missing in the STC visualizations.
Nevertheless, future works should explore how to adapt the projection operation to extract more reliable information. For example, the clipping plane could be adjusted over time to track an object of interest or take into account the internal motions of the objects, potentially extending the cutting plane to an arbitrary surface. 

A second aspect that should be further studied, is the usage of the STH for other spatio-temporal datasets.
The examples and use cases described in this paper came from the biology domain, and in particular on the analysis of morphogenesis. 
In the explored datasets, the temporal and spatial coherency was high, i.e. motions between two consecutive data points were relatively low, with quite a low spatial density, and the different structures visualized are well-defined (i.e. cells). 
%
%The STC visualizations had a number of cavities which allowed a good tracking of the cells and were easily explored using the additional cutting operations. 
%
It remains unknown whether the STH approach would be still usable for spatio-temporal datasets with lower temporal and spatial coherency.%, and higher spatial density. 
In such situations, the STC visualization could be denser and less continuous, potentially impacting its comprehension. 

Finally, a recurrent comment from the domain experts was the addition of annotation functionalities to the system.
In this respect, in the morphogenesis context, domain experts considered that the tool could be of interest for data curation, in particular to validate/complete the object tracking algorithms over time, notably after cell divisions, or to label cells.
Such operations can be very cumbersome on desktop applications due to the multidimensional nature of the data.
The STC visualizations could help for detecting errors or areas of interests and use them to directly annotate those events.

\section{Conclusion}\label{conclusion} 

In this paper, we have proposed a novel spatio-temporal visualization based on the Space-Time Cube visualization. The proposed visualization, the Space-Time Hypercube, extends the STC visualization to consider a third spatial dimension in the data.
%
%% Talk about the operation in 4D
However, the interaction and manipulation of 4D remains unpractical. Thus, to enable a direct visualization, we proposed a projection operator on based on a user-driven cross-section defined in 3D space.
%
%% Loss of information
The projection of the hypercube, consequently a STC, contains only a partial spatial information of the dataset, but creates a view containing temporal information. 
Numerical and categorical information could be displayed as well on the visualization.
%
%%VR framework
We juxtaposed and linked the STC and original dataset visualizations in a VR application, taking advantage of immersive environments benefits in terms of visualization and interaction in 3D.
Various tools for exploration, filtering or tracking objects apply transformations on both visualization and come to assist in the analysis of the dataset.
%
%%Evaluation
Moreover, we illustrated the potential usages of the STC in several use cases in the context of morphogenesis. 
The evaluation of the STH in a context of embryology showed that the STC visualization presents a number of benefits with respect to traditional visualizations, specially for the detection of events with a temporal relevance. 
The STH approach could pave the way to new types of visualization and interaction methods for 3D spatio-temporal data, and we believe that such tools will help the adoption of VR technologies for data visualization.

%%keep the beginning, let's just add a generalization consideration or something, like this paragraph before ending

%Retrospectively, we believe that STH is a opening to new types of visualization and interaction. As the generation algorithm was made to adapt to any type of data encoding, we plan to extend this method to other types of 3D temporal dataset. 
%
%Exploring meaningful operations of projection and interaction on the 4D data could notably help in visualizing other particular 3D temporal objects. 
%
%As Section \ref{UCdeformation} implied, our projection can help in visualizing structural evolution and local movements, thus we suggest that other projections could focus on object movements and trajectories.

%To conclude, we believe that creating relevant visualizations in immersive environment will entice analysts to consider such technology in their workflow.
%retrospective stuff
%Our method is independent from data type, so we hope this could be used for a large range of datasets


\bibliographystyle{splncs04}
\bibliography{biblio}


%

%Author, F.: Article title. Journal \textbf{2}(5), 99--110 (2016)

%Author, F., Author, S.: Title of a proceedings paper. In: Editor,
%F., Editor, S. (eds.) CONFERENCE 2016, LNCS, vol. 9999, pp. 1--13.
%Springer, Heidelberg (2016). \doi{10.10007/1234567890}

%Author, F., Author, S., Author, T.: Book title. 2nd edn. Publisher,

%Author, A.-B.: Contribution title. In: 9th International Proceedings
%on Proceedings, pp. 1--2. Publisher, Location (2010)

%LNCS Homepage, \url{http://www.springer.com/lncs}. Last accessed 4
%\end{thebibliography}
\end{document}